\documentclass[preprint,showpacs,preprintnumbers,amsmath,amssymb]{revtex4}

\usepackage{graphicx}
\usepackage{bm}

\begin{document}

\title{A formalism for studying long-range correlations in many-alphabets sequences.}

\author{S. L. Narasimhan$^*$, Joseph A. Nathan$^+$, P. S. R. Krishna and K. P. N. Murthy$^{**}$ }
\affiliation{Solid State Physics Division, $^+$Reactor Physics Design Division\\
             Bhabha Atomic Research Centre, Mumbai-400085, India.\\
$^{**}$Materials Science Division, Indira Gandhi Centre for Atomic Research,\\
Kalpakkam 603102, Tamilnadu, India.}
\email{slnoo@magnum.barc.ernet.in}

\vspace{3cm}

\begin{abstract}
We formulate a mean-field-like theory of long-range correlated $L$-alphabets
sequences, which are actually systems with $(L-1)$ independent parameters. Depending on the values of these parameters, the variance on the average number of any given symbol in the sequence shows a linear or a superlinear dependence on the total length of the sequence. We 
present exact solution to the four-alphabets and three-alphabets sequences. We also demonstrate 
that a mapping of the given sequence into a smaller alphabets sequence (namely, a {\it coarse-
graining} process) does not necessarily imply that long-range correlations found in the latter would correspond to those of the former. 
\end{abstract}

\maketitle

\section{Introduction} A standard method for studying the stochastic behavior of complex physical, chemical or even biological systems consists first in dividing the set of all states available to them into a {\it finite} number of distinct classes, labelled by distinct symbols, and then in representing the dynamical evolution of these systems as sequences of these symbols [1,2,3]. Statistical correlation between the symbols of this representative sequence, if any, will then tell us about the way the original system evolves. In particular, the presence of long-range correlations(LRC) in the symbolic sequence will suggest a history-dependence of the system's evolution. 

If on the other hand these sequences represent the states available to the various parts of the system under study, then their statistical properties could also throw light on the way these subsystems are organized. For example, natural language texts are made up of words put together according to certain syntactic rules; so, when they are treated as sequences of alphabets, their syntactic structure and semantic content should manifest as correlations between the alphabets. Since the rules for putting words together do not extend beyond a sentence, we may expect the syntactic structure of the text to show up as short-range correlations, whereas we may expect the semantic content of the text to show up as long-range correlations between the alphabets.

Conversely, it is of interest to know whether long-range correlations present in the representative symbolic sequence always implies a {\it non-local or global} behavior of the original system and if so, whether they can also provide a parametric description. Reducing the number of alphabets or symbols, called {\it coarse-graining} of the sequence, is equivalent to reducing the number of parameters in the problem; such a reduction of parameters is not expected to affect long-range correlations that might be present in the sequence but may however lead to a loss of short-range correlations. Therefore, if the interest is in studying the {\it non-local or global} behavior of the system that shows up as long-range coorelation between symbols of its  representative sequence, then we may as well group the symbols into two distinct classes and represent the system in terms of a binary sequence. 

Recently, it has been suggested [4] that non-trivial correlations in a binary sequence could be due to the presence of a long-range memory whose strength exceeds a certain critical value. More specifically, the number of zeros, $N_0$, in given a binary sequence $S_N \equiv \{ \sigma _i = 0,1 \mid i = 1,\dots,N\}$, is expected to be close to $N/2$ if the sequence $S_N$ were long and unbiassed. It is, in fact, a random variable gaussian-distributed around its expected mean value, $N/2$, with a variance denoted by $\sigma ^2 (N)$. It has been shown by Keshet and Hod [4] that the existence of non-trivial correlations in the sequence can be inferred from the anomalous power-law dependence of $\sigma ^2 (N)$ on $N$ -  namely, $\sigma ^2 (N) \sim N^{2\mu}$ where $\mu$, the strength of long-range memory, exceeds the critical value $1/2$. This mean-field-like theory of correlated binary sequences seems to provide a paradigm for studying the correlational properties of generic symbolic sequences such as DNA and even natural language texts. 

The implicit assumption in this approach is that the many-alphabets sequence representing an LRC system can always be coarse-grained into a binary sequence without losing the long-range correlational properties of the original sequence. There is no {\it a priori} reason why this assumption should be expected to hold good in general. For example, it has been argued [5] that a minimum of ten letters (not less than five letters, in any case) are needed for designing a foldable model of amino acid (twenty-alphabets) sequences. This example suggests that the minimum number of symbols required for a sequential representation of a system (or equivalently, the extent to which the state-space of a system can be coarse-grained) may depend on the specific behaviour of the system under study. 

In the next section, we present a mean-field-like theory of an $L$-alphabets sequence with long-range memory, which is a generalization of our earlier study on ternary sequences [6]. In the third section, we work out the exact phase diagrams for the special cases of four-symbol and three-symbol problems. We also show that a mapping of the ternary sequence to the binary sequence could lead to spurious correlations. Whether this result holds good for a generic symbolic sequence is a moot question because the formulation presented here deals with non-stationary sequences. We summarize the results in the last section. 

\section{Many-alphabets sequences} In order to study the statistical properties of such sequences, we first need to define the conditional probability, $p(t_i \in \alpha \mid T_{N,i})$, that the $i$th symbol, $t_i$, in the sequence will be one of $\alpha =\{ 0,1,2,...,L-1\}$, where $T_{N,i} (\equiv t_{i-N} t_{i-N+1},...,t_{i-1})$ denotes the earlier part of the sequence under consideration. Depending on whether $T_{N,i}$ denotes only a part or the whole of the earlier sequence, we may say that the sequence being studied has {\it finite} or {\it infinite memory}. We are concerned here with sequence with {\it infinite or unbounded} memory. The conditional probability, $p(t_i \in \alpha \mid T_{N,i})$, could in general depend not only on the individual symbols but also on their specific positional ordering in the sequence. Ignoring the possible configuration-dependence of this conditional probability leads to a solvable mean-field-like theory of these sequences.

We first define the conditional probability, $p(t_{i}=0\mid T_{N,i})$, of finding zero as the $i^{th}$ symbol in the sequence as follows:
\begin{equation}
p(t_i=0 \mid T_{N,i}) = \frac{1}{N} \sum \limits _{j=0}^{L-1} n_j g_j
\end{equation}
where $n_j$'s denote the number of $j$'s in the sequence such that $\sum _{j\in \alpha} n_j =N$, and $g_j$'s denote the {\it a priori} probabilities of choosing the respective symbols; only $L-1$ of the symbols are {\em independent} because the {\it a priori} probabilities for the symbols will have to add up to unity ({\it i.e.,}$\sum _{j\in \alpha} g_j =1$).

The conditional probabilities, $p(t_i =j\mid T_{N,i})$, for the rest of the symbols, $j = 1,2,...,L-1$, to be found at the $i^{th}$ position of the sequence may be defined [6] in terms of the complementary sequences $T^j _{N,i} (\equiv t^j _{i-N}t^j _{i-N+1},...,t^j _{i-1})$, where $t^j _{i-k} \equiv (t_{i-k}+j)($ mod $L)$ for $k = 1,2,...,N$:
\begin{eqnarray}
p(t_i =j\mid T_{N,i}) & \equiv & p(t_i =0\mid T^j _{N,i}) \nonumber\\
                       & = & \frac{1}{N} \sum \limits _{l=0}^{L-1} n_l g_{(j+l)(\mod L)} \\
                       & = & \frac{1}{N} \sum \limits _{l=0}^{L-1} n_{(L-j+l)(\mod L)} g_l
\end{eqnarray}
Clearly, these definitions ensure that $\sum _j p(t_{i}=j\mid T_{N,i}) = 1$.
We may now parametrize the deviations of $g_l$ from their 'unbiassed' values $1/L$ by writing $g_l = (1+\mu _l)/L$ where $\mu _{l\in \alpha } \in [-1,L-1]$ and $\sum _{l\in \alpha} \mu _l = 0$, and then
reexpress the conditional probabilities in terms of the independent symbols, $\{ 0,1,2,...,L-2\}$: 
\begin{eqnarray}
p(t_i =0\mid T_{N,i}) & = & \frac{1}{L} \left( 1 + \frac{1}{N} 
 \sum \limits _{l=0}^{L-2} {\cal N}_l \mu _l\right)\\
\nonumber \\
p(t_i =j>0\mid T_{N,i}) & = & \frac{1}{L} \left( 1 - \frac{1}{N}{\cal N}_{L-j-1} \mu _{j-1} + \frac{1}{N} \sum \limits _{l=0(\neq j-1)}^{L-2} {\tilde n}_l \mu _l \right) \\
{\cal N}_m & \equiv & \left( 2n_m + \sum \limits _{k=0(\neq m)}^{L-2} n_k \right) - N \\
{\tilde n}_l & \equiv & n_{(L-j+l)(\mod L)}-n_{(L-j-1)}                                                
\end{eqnarray}
We may also denote these conditional probabilities by $p_j (n_0 , n_1 ,\cdots ,n_{L-2};N)$
in order to show the independent symbols and variables explicitly. We may simplify the 
notations further by defining ${\bf n} \equiv (n_0 ,n_1 ,\cdots ,n_{L-2})$ and 
${\bf n}_j -1 \equiv (n_0 ,n_1 ,\cdots ,n_j -1,\cdots ,n_{L-2})$. Accordingly,
$p(t_i = j\mid T_{N,i})$ could be denoted by $p_j ({\bf n};N)$.

Now, a sequence of $N$ symbols is completely described by the probability, $Q({\bf n};N)$, that there are $n_{0}$ zeros, $n_{1}$ ones and so on in the sequence. We may write the following discrete equation for $Q({\bf n};N)$:
\begin{equation}
Q({\bf n};N+1) = \sum \limits _{j=0}^{L-2}p_j ({\bf n}_j -1;N)Q({\bf n}_j -1;N) + 
\left(1- \sum \limits _{j=0}^{L-2}p_j({\bf n};N)\right)Q({\bf n};N)
\end{equation}
Since we expect the average number of any symbol in the sequence to be close to its asymptotic value, $N/L$, we may rewrite the above equation in terms of the variables, $x_j \equiv Ln_j -N$ for all $j = 0,1,\cdots ,L-2$. In doing so, we make use of the correspondence, 
\begin{eqnarray}
({\bf n}; N+1) & \rightarrow & (x_0 -1,x_1 -1,\cdots ,x_{L-2}-1; N+1) 
\equiv ({\bf X}-1;N+1)\nonumber \\
({\bf n}_j -1 ;N) & \rightarrow & 
(x_0,x_1 ,\cdots ,x_j -L,\cdots ,x_{L-2} ; N)\quad \, \, \, \equiv ({\bf X}_j -L;N)\nonumber 
\end{eqnarray}
and rewrite Eq.(8) in the form, 
\begin{eqnarray}
Q({\bf X}-1; N+1 ) = \sum \limits _{j=0}^{L-2} p_j ({\bf X}_j -L ;N)
Q({\bf X}_j -L ;N) + \left(1- \sum \limits _{j=0}^{L-2}p_j({\bf X};N)\right)Q({\bf X};N)
\end{eqnarray}
The transition probabilities, $p_j({\bf X},N)$ are given by
\begin{equation}
p_j ({\bf X};N) \equiv \frac{1}{L}\left( 1 + \frac{1}{N}f_j ({\bf X},\lambda)\right) ; 
                                                         \quad j = 0,1,\cdots L-2
\end{equation}
where the 'drift' forces, $f$'s are defined as follows:
\begin{eqnarray}
f_0 ({\bf X},\lambda) & = & \sum _{l=0}^{L-2} \lambda _l x_l \\
f_1 ({\bf X},\lambda) & = & -\lambda _0 x_{L-2}
                            +\sum _{l=0}^{L-3}(\lambda _{l+1} -\lambda _0 )x_l \\
f_{j\geq 2} ({\bf X},\lambda) & = & -\lambda _{j-1} x_{L-j-1} 
                        +\sum _{l=0}^{L-j-2}(\lambda _{l+j} -\lambda _{j-1} )x_l 
                        +\sum _{l=L-j}^{L-2}(\lambda _{l+j-L} -\lambda _{j-1} )x_l \\
\lambda _m & = & \frac{1}{L}\left( 2\mu _m + \sum \limits _{k=0(\neq m)}^{L-2} \mu _k \right)
\end{eqnarray}
For a long chain ($N>>1$), the hopping process described by Eq.(9) becomes more transparent in the continuum limit:
\begin{equation}
\frac{\partial Q({\bf X};N)}{\partial N} = \frac{1}{2}(L-1)\sum \limits _{j=0}^{L-2}
\left( \frac{\partial ^2 Q({\bf X};N)}{\partial x_j^2} - 
\frac{1}{N+N_{0}} \frac{\partial [f_j (\bf X,\lambda)Q({\bf X};N)]}{\partial x_j}\right)
\end{equation}
The parameter $N_{0}$ has been introduced as a cutoff below which the above continuum version of Eq.(9) would not be meaningful. It could depend on the memory parameters, $\mu$'s [4].

In order to solve this equation subject to the initial condition, $Q({\bf X} ;N=0)=\delta (\bf X)$, we first Fourier transform it with respect to the variables $x_j$'s:
\begin{equation}
\frac{\partial {\tilde Q({\bf q} ;N)}}{\partial N} = 
\sum \limits _{j=0}^{L-2}\left( -\frac{1}{2} (L-1)q_j^2 {\tilde Q({\bf q} ;N)} 
+ \frac{1}{N+N_{0}}f_j ({\bf q};\lambda) \frac{\partial {\tilde Q({\bf q} ;N)}}{\partial q_j} 
\right)
\end{equation}
Here, $q_j$'s are the Fourier conjugates of the $x_j$'s and the $f_j ({\bf q},\lambda)$'s
are obtained from Eqs.(11-14) by substituting $q_j$'s for $x_j$'s. 
This first order equation can then be solved by the method of characteristics. In particular, we have to solve the equations,
\begin{equation}
\gamma\frac{dN}{N+N_{0}} = \frac{dq_0}{f_0 ({\bf q},\lambda)}                                         
                         = \frac{dq_1}{f_1 ({\bf q},\lambda)}
                         = \cdots = \frac{dq_{L-2}}{f_{L-2} ({\bf q},\lambda)}
\end{equation}
The standard method of solving them leads to rewriting Eq.(16) in terms of the {\it normal} coordinates, $\tilde q^{(l)}$'s [7]:
\begin{equation}
\frac{\partial {\tilde Q(\tilde {\bf q} ;N)}}{\partial N} = 
\sum \limits _{l=0}^{L-2}\left( 
                        -\frac{1}{2}(L-1)(\tilde q^{(l)})^2 {\tilde Q(\tilde {\bf q} ;N)} 
+ \frac{1}{N+N_{0}}\rho ^{(l)} \tilde q^{(l)} 
                         \frac{\partial {\tilde Q(\tilde {\bf q} ;N)}}{\partial \tilde q_j} 
\right)
\end{equation}
where the $\rho$'s are the eigenvalues of the 'LRC' matrix,
\begin{equation}
\Lambda _L \equiv \left( \begin{array}{cccccc}
        \lambda _0 & \lambda _1 & \lambda _2 & \cdots & \lambda _{L-3} & \lambda _{L-2} \\
        \lambda _1 -\lambda _0 & \lambda _2 -\lambda _0 & \lambda _3 -\lambda _0 & \cdots &  \lambda _{L-2} -\lambda _0 & -\lambda _0 \\
        \lambda _2 -\lambda _1 & \lambda _3 -\lambda _1 & \lambda _4 -\lambda _1 & \cdots & -\lambda _1 & \lambda _0 -\lambda _1 \\
        \cdots & \cdots & \cdots & \cdots & \cdots & \cdots \\
        \lambda _{L-3} -\lambda _{L-4} & \lambda _{L-2} -\lambda _{L-4} & -\lambda _{L-4} 
                & \cdots & \lambda _{L-6} -\lambda _{L-4} & 
                                                   \lambda _{L-5} -\lambda _{L-4} \\
        \lambda _{L-2} -\lambda _{L-3} & -\lambda _{L-3} & \lambda _0 -\lambda _{L-3} 
                & \cdots & \lambda _{L-5} -\lambda _{L-3} & 
                                                   \lambda _{L-4} -\lambda _{L-3} \\
       \end{array} \right)
\end{equation}
A further transformation,
\begin{equation}
\xi ^{(l)} \equiv \tilde q^{(l)} \tau ^{\rho ^{(l)}}; \quad \tau \equiv N + N_0 ;
\quad l = 0,1,\cdots , L-2
\end{equation}
leads to the simpler form,
\begin{equation}
\frac{\partial {\tilde Q({\bf \xi} ;\tau )}}{\partial \tau} = 
-\frac{1}{2}(L-1)\left( 
\sum \limits _{l=0}^{L-2} (\xi ^{(l)})^2 \tau ^{-2\rho ^{(l)}} 
\right) \tilde Q({\bf \xi} ;\tau)
\end{equation}
whose solution can be written down immediately. Transforming back to the variables, $\tilde q^{(l)}$ and using the initial condition, $\tilde Q(\tilde {\bf q} ;\tau = N_0) = 1$, we have,
\begin{equation}
\tilde Q(\tilde {\bf q} ;\tau) = \exp \left \{ -\frac{1}{2}(L-1)
\left( \sum \limits _{l=0}^{L-2} 
\left[ 1 - \left( \frac{N_0}{N+N_0} \right)^{-2\rho ^{(l)}+1} \right] 
\frac{(\tilde q^{(l)})^2}{(1 - 2\rho ^{(l)})}\right) \tau \right \}
\end{equation}
Since the maximum eigenvalue, $\rho \equiv $ max ${\rho ^{(l)}}$, decides the asymptotic behaviour of $\tilde Q$, the variance associated with the number of any given symbol turns out to be
\begin{equation}
\sigma ^2 (\tau) = \left[ 1 - \left( \frac{N_0}{N+N_0} \right)^{-2\rho +1} \right]
\frac{\tau}{(1-2\rho )} ;\quad 2\rho \neq 1
\end{equation} 
It is clear from the above expression that $\sigma ^{2}(\tau) \propto \tau ^{\nu}$,
where $\nu =1$ whenever $2\rho < 1$ and $\nu =2\rho$ whenever
$2\rho > 1$. Even in the case, $2\rho = 1$, the exponent $\nu =1$
but there will be logarithmic corrections.
The existence of a critical value, $\rho _c = 1/2$, for $\rho$ implies that the $(L-1)$ dimensional parameter space, $\{\mu _l \mid -1\leq \mu _l \leq (L-1); l = 0,1, \cdots , L-2\}$ is divided into two distinct regions, diffusive ($\nu = 1$) and superdiffusive ($\nu > 1$). In the next section, we present exact solutions to the special cases, ternary and quaternary sequences.

\section{Special cases} \noindent A. {\it Quaternary (four-symbols) sequence}:

There are only three independent correlation parameters, $\mu _0 ,\mu _1, \mu _2 \in [-1,3]$, for a quaternary sequence which define the $\lambda$'s, Eq.(14):
\begin{eqnarray}
\lambda _0 & = & \frac{1}{4}\left( 2\mu _0 + \mu _1 + \mu _2 \right) \nonumber \\
\lambda _1 & = & \frac{1}{4}\left( \mu _0 + 2\mu _1 + \mu _2 \right) \\
\lambda _2 & = & \frac{1}{4}\left( \mu _0 + \mu _1 + 2\mu _2 \right) \nonumber
\end{eqnarray}
The corresponding LRC matrix is then given by Eq.(19),
\begin{equation}
\Lambda _4 \equiv \left( \begin{array}{ccc}
        \lambda _0 & \lambda _1 & \lambda _2 \\
        \lambda _1 -\lambda _0 & \lambda _2 -\lambda _0 & -\lambda _0 \\
        \lambda _2 -\lambda _1 & -\lambda _1 & \lambda _0 -\lambda _1 
       \end{array} \right)
\end{equation}
whose eigenvalues are given by
\begin{eqnarray}
\rho _0 & = & \lambda _0 -\lambda _1 +\lambda _2 = \frac{1}{2}(\mu _0 + \mu _2) \nonumber \\
\rho _1 & = & \sqrt{(\lambda _0 -\lambda _2)^2 + \lambda _1^2}  = 
              \sqrt{\frac{1}{8}\left[ (\mu _0 +\mu _1 )^2 + (\mu _1 + \mu _2 )^2 \right]} \\
\rho _2 & = & -\rho _1 \nonumber
\end{eqnarray}
Since $\rho _1$ is always positive, the maximum value between $\rho _0$ and $\rho _1$ decides which part of the $(\mu _0 ,\mu _1 ,\mu _2)$-parameter space is diffusive and which part is super-diffusive. 
%
%
\begin{figure}
\includegraphics[width=3.25in,height=2.75in]{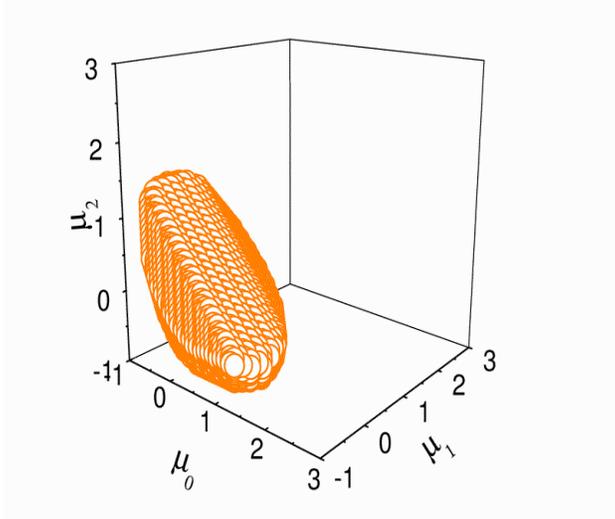}
\caption{The diffusive part of the phase space associated with the quaternary sequence. It is slanted and ellipsoidal in shape, cut by the coordinate planes as well as by the plane $\mu _0 + \mu _1 + \mu _2 = 1$.}
\end{figure}
%
%
%
%
\begin{figure}
\includegraphics[width=3.25in,height=2.75in]{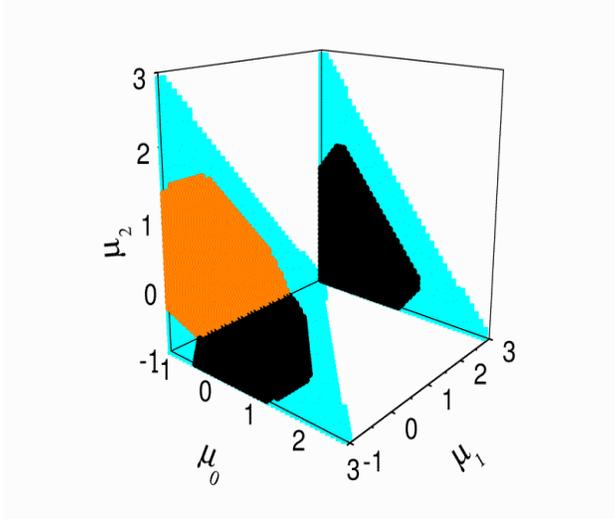}
\caption{The projections of the diffusive region shown in Fig.1 on the three coordinate planes. The triangular regions shown in lighter hue are the projections of the available phase space prism containing the diffusive region shown in Fig.1.}
\end{figure}
The diffusive part of the available phase space is shown in Fig.1. It is slanted and ellipsoidal in shape, cut by the bounding planes. Its projections on the $(\mu _0,\mu _1), (\mu _1,\mu _2)$ and $(\mu _0,\mu _2)$ planes are shown in Fig.2. Also shown in the figure are the projections of the available phase space containing the truncated ellipsoidally shaped diffusive region. 

\noindent B. {\it Ternary (three-symbols) sequence} [6]:

There are only two independent parameters, $\mu _0 ,\mu _1 \in [-1,2]$, for the ternary sequence and the corresponding LRC matrix, 
\begin{equation}
\Lambda _3 \equiv \left( \begin{array}{cc}
        \lambda _0 & \lambda _1 \\
        \lambda _1 -\lambda _0 & -\lambda _0 
       \end{array} \right)
\end{equation}
has a positive eigenvalue given by
\begin{equation}
\rho = \sqrt{\lambda _0^2 +\lambda _1^2 -\lambda _0 \lambda _1} = 
       \sqrt{\frac{1}{3}(\mu _0^2 +\mu _1^2 +\mu _0 \mu _1)}
\end{equation}
where $\lambda _0 = (2\mu _0 +\mu _1)/3$ and $\lambda _1 = (\mu _0 + 2\mu _1)/3$. The diffusive region of the $(\mu _0,\mu _1)$-parameter space is then defined by the condition $\rho \leq 1/2$ and corresponds to the elliptical region shown in Fig.3. 
%
%
\begin{figure}
\includegraphics[width=3.25in,height=2.75in]{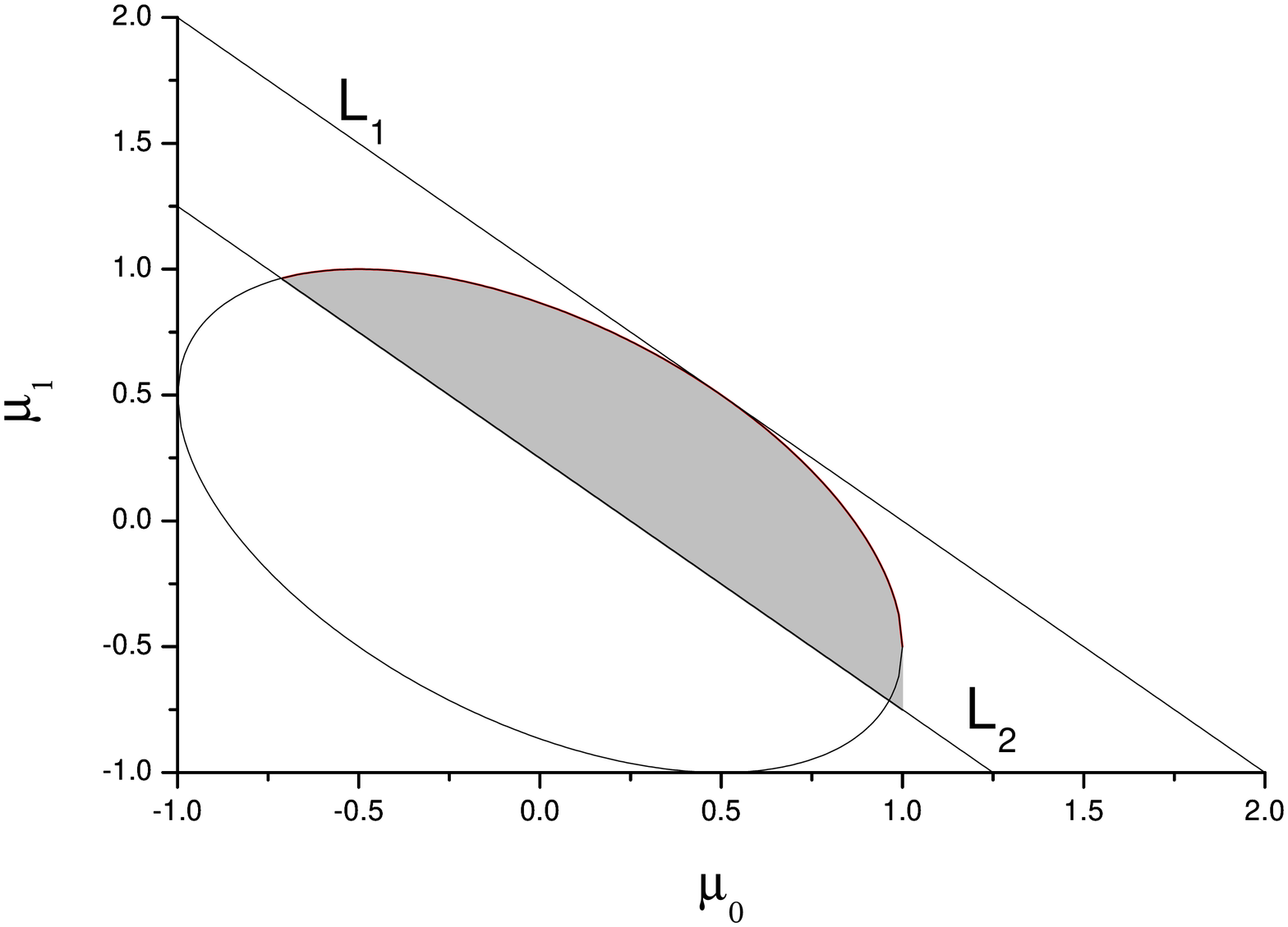}
\caption{Phase diagram for the ternary sequence. $L_1$ and $L_2$ refer to the lines $\mu _0 + \mu _1 = 1$ and $\mu _0 + \mu _1 = 1/4$ respectively. The ellptical region of the triangular phase space corresponds to the {\it diffusive} behavior whereas the region exterior to it corresponds to the 
{\it superdiffusive} behavior of the ternary sequences. The entire region between the lines $L_1$ and $L_2$, inclusive of the shaded elliptical region, gets mapped into the superdiffusive regime of the binary.}
\end{figure}
\noindent The area of this diffusive region is $\pi \sqrt{3}/2$ which is roughly $60\%$ of the available phase space area. This may be contrasted with the binary case where it is exactly $50\%$. Now the question arises whether a diffusive subregion of the ternary is likely to be mapped into a superdiffusive region of the binary due to a coarse-graining process.
Since a mapping of a set of three symbols to a set of two symbols always introduces a global bias in the system, we will have to reformulate the binary case accordingly.

\noindent C.   {\it Mapping the ternary into the binary}:

The {\it a priori} probabilities assigned to the symbols of the ternary sequence are
\begin{equation}
g_i = \frac{1}{3}(1 + \mu _i);\quad \mu _0 +\mu _1 +\mu _2 = 0; \quad i = 0,1,2 
\end{equation}
where the memory parameters, $\mu _i$'s, have their values in the range, $[-1, 2]$ and satisfy  the above constraint. If we replace, for example, the symbol one of the ternary by symbol zero, we have a binary sequence of symbols, zero and one, with the {\it a priori} probabilities,
\begin{equation}
g_1 = \frac{1}{3} (1 + \mu _2 );\quad g_0 = \frac{1}{3} (2 + \mu _0 + \mu _1) = \frac{2}{3}(1-\frac{\mu _2}{2})
\end{equation}
where the constraint given in Eq.(29)has been made use of. The above definitions are of the general form,
\begin{equation}
g_1 = b(1 +\mu _2);\quad g_0 = b^{'}\left( 1 - \frac{b}{b^{'}}~\mu _2\right)
\end{equation}
with $b = 1/3$ and $b^{'} \equiv (1 - b) = 2/3$. This implies that a ternary sequence without memory ($\mu _0 =\mu _1 =\mu _2 = 0$ in Eq.(29)) corresponds to a binary sequence without memory ($\mu _2 = 0$ in Eq.(30)) but consisting of unequal number of zeros and ones. This is at variance with what is implied by Eq.(1) which provides the simplest mechanism for having a binary sequence with unbounded memory. 

In fact, Eq.(1) defines the conditional probability, $p_0 (n_0 ,N)$, of finding zero as the $(N+1)^{th}$ symbol of the sequence that already has $n_0$ zeros:
\begin{equation}
p_0 (n_0 ,N)  \equiv  \frac{1}{N} \left( n_0 g_0 + n_1 g_1 \right) 
= \frac{1}{2}\left( 1 + \frac{(2g_0 - 1)}{N}[2n_0 - N]\right)
\end{equation}
It is clear from this definition that $(2g_0 - 1)$ parametrizes the memory-dependence for the occurrence of a symbol in the sequence. A sequence without memory corresponds to the unbiassed case ($g_0 = 1/2 = g_1$) and, on the average, has equal number of zeros and ones. 

In general, for arbitrary values of $g_0$, the mean difference between the number of zeros and the number of ones has recently been shown [8] to have an asymptotic form,
\begin{equation}
<\delta n> \equiv <\mid n_0 - n_1\mid > \sim \frac{2q-1}{\Gamma (2g_0)}N^{2g_0-1}
\end{equation}
where $q$ is the {\it a priori} probability of having the first symbol of the sequence as zero. For $g_0 < 1/2$, the mean difference, $<\delta n>$, vanishes even if the first symbol were chosen with a biassed coin. In other words, the average number of zeros tends to be equal to the average number of ones for $g_0 < 1/2$.

This implies that Eq.(32) is not the right definition for $p_0(n_0, N)$ because we expect the average number of zeros, $<n_0>$, in the sequence to be equal to $b^{'}N$ for $\mu _2 > (1-2b)/2b$ (equivalently, for $g_0 < 1/2$). Since the natural variable for a Markov chain generated with a biassed coin is $x (~\equiv [n_0/b^{'} - N]$, say), it is likely that the conditional probability $p_0(n_0, N)$ depends on $n_0$ through the variable $x$. We therefore try the {\it ansatz}, 
\begin{eqnarray}
p_0 (x,N) & = & \alpha + \frac{\beta}{N}~x \\
\beta     & \equiv & b^{'}(2g_0 - 1) 
\end{eqnarray}
where $\alpha$ is a parameter yet to be fixed. The probability, $Q(x,N)$, that the number of zeros in a chain of length $N$ is $n_0$ can now be shown to satisfy the following equation in the continuum limit:
\begin{equation}
\frac{\partial Q}{\partial N} = \frac{1}{2b^{'2}}(1 - b^{'2})\frac{\partial ^2 Q}{\partial x^2}
+ (1 - \frac{\alpha}{b^{'}})\frac{\partial Q}{\partial x} 
- \frac{\beta}{b^{'}(N + N_0)}\frac{\partial [xQ]}{\partial x}
\end{equation}
where, as mentioned earlier, $N_0$ denotes a cutoff value for $N$ below which the above continuum description is not valid.

This equation suggests that an interesting description from the LRC's point of view corresponds to the case when $\alpha = b^{'}$ [9] leading to the following general definition for $p_0(n_0, N)$:
\begin{equation}
p_0(n_0, N) \equiv b^{'}\left(1 + \frac{(2g_0 - 1)}{N}\left[\frac{n_0}{b^{'}} - N\right] \right)
\end{equation}
which reduces to that given by Eq.(32) for $b^{'} = 1/2$. The first moment of the distribution, $<x>$, then turns out to have the interesting form, $<x> \sim (N + N_0)^{2g_0-1}$, while the variance of the distribution is given by,
\begin{equation}
\sigma ^2 (N) \propto \left\{\begin{array}{lll}
                              N & \mbox{for} & (2g_0 - 1)\leq 1/2 \\
                              N^{2(2g_0 - 1)} & \mbox{for} & (2g_0 - 1) > 1/2
                             \end{array}
                      \right.
\end{equation}
In other words, long-range correlations in a binary sequence are characterized by the exponent, $(2g_0 - 1)$, for $g_0 > 3/4$ which in turn corresponds to having the values of $\mu _2$ in the range $\mu _2 \in (-1, -1 + [1/4b])$ by Eq.(31).

Since $\mu _2 = -(\mu _0 + \mu _1)$ and $b = 1/3$, the condition for non-trivial correlations in the coarse-grained binary sequence turns out to be $\mu _0 + \mu _1 > 1/4$, above
the lower line shown in Fig.3. The entire region of the phase space between this line and the
upper line, $\mu _0 + \mu _1 = 1$, corresponds to long-range correlation when mapped into the biassed binary.

Defining the memory parameter, $\mu $, for the binary sequence simply as $\mu \equiv -\mu _2/2$, we see that the phase space region, $1/4 \leq (\mu _0 + \mu _1) \leq 1$, of the ternary sequence corresponds to the parameter range, $1/8 \leq \mu \leq 1/2$, of the biassed binary sequence; every line parallel to and in between the upper and lower lines in Fig.3 is mapped into a point in the range $1/8 \leq \mu \leq 1/2$ and {\it vice versa}. In particular, the shaded part of the ellipse in Fig.3 is the diffusive area that is mapped into the superdiffusive regime of the biassed binary, which in general is characterized by a parameter, $\mu$, in the range $(-1 + [3/4b], -1 + [1/b])$.

Conversely, the fact that the correlation parameter of the biassed binary sequence under study has a value in the range $\mu \in (-1 + [3/4b], -1 + [1/b])$ does not necessarily mean that the parent ternary sequence also has long-range correlations. Even if we know {\it a priori} that there are long-range correlations between symbols of the parent sequence, the parameter $\mu$ is not a true measure of its strength.

\section{Summary}

We have presented a formalism for studying the correlation properties of a many-symbols sequence and, by way of illustration, obtained the phase diagrams for quaternary and ternary sequences. Motivated by the fact that the diffusive portion of the phase space is larger for the ternary than for the binary, we have studied a mapping between the two. It turns out that long-range correlation for the binary does not necessarily imply long-range correlation for the ternary. This result has deeper implications for the coarse-graining of the many-alphabets sequence. For example, if we do not know {\it a priori} that there are long-range correlations in the original sequence, then we cannot conclude that long-range correlation found in a coarse-grained sequence implies that in the original sequence. On the other hand, even if we know {\it a priori} that there are long-range correlations in the original sequence, we cannot conclude that the correlation parameters of the coarse-grained sequence represent the true strength of correlations in the parent sequence, if we do not know how the coarse-grained sequence had been obtained. It is also likely that different coarse-graining {\it paths} from the parent sequence to lower dimensional representative sequence could lead to different inferences [10]. The applicability of this result to a generic symbolic sequence needs further investigation since it is obtained in a formalism dealing with non-stationary sequences.

SLN wishes to acknowledge helpful discussions with Y. S. Mayya.

\vspace{0.3cm}
\noindent
{\bf REFERENCES}
\begin{enumerate}

\item U. Balucani, M. H. Lee and V. Tognetti, Phys. Pep. {\bf 373}, 409 (2003).

\item H. E. Stanley {\it et al}, Physica, A{\bf 224},302 (1996); R. N. Mantegna and H. E. Stanley, Nature (London) {\bf 376}, 46 (1995); A. Provota and Y. Almirantis, Physica, A{\bf 247}, 482 (1997), and references therein.

\item I. Kanter and D. A. Kessler, Phys. Rev. Lett. {\bf 74}, 4559 (1995); P. Kokol, J. Brest and J. Zumer, Cybernetics and systems, {\bf 28}(1), 43 (1997); W. Ebeling and T. Poschel, 
Europhys. Lett. {\bf 26}, 241 (1994), and references therein.

\item Shahar Hod and Uri Keshet, Phys. Rev. E{\bf 70}, R015104 (2004) ; O. V. Usatenko and V. A. Yampol'skii, Phys. Rev. Lett. {\bf 90}, 110601 (2003);
O. V. Usatenko and V. A. Yampol'skii, arXiv:physics/0211053.

\item K. Fan and W. Wang, J. Mol. Biol. {\bf 328}, 921 (2003).

\item S. L. Narasimhan, Joseph A. Nathan and K. P. N. Murthy, Europhys. Lett. {\bf 69}, 22 (2004)

\item M. C. Wang and G. E. Uhlenbeck, Rev. Mod. Physics, {\bf 17}, 323 (1945)

\item G. M. Schutz and S. Trimper, Phys. Rev. E{\bf 70}, R045101 (2004)

\item In the case when $\alpha \neq b^{'}$, the first moment $<x>$ is asymptotically of the form $(\alpha - b^{'})N/b^{'}(1-g_0)$ implying thereby that the average number of zeros is never equal to $b^{'}N$.

\item M. Kak and J. Logan, in {\it Fluctuations Phenomena (Studies in Statistical Physics)}
Ed. by E. W. Montroll and J. L. Lebowitz, North-Holland Publishing company (1979), page 10,
point out that a coarse-grained description may not preserve the Markovian character of the
process under study.

\end{enumerate}

\end{document}